\documentclass[prl,twocolumn,showpacs]{revtex4}
\usepackage{psfig}
\usepackage{amssymb}
\begin{document}
\def\/{\over}
\def\({\left(}
\def\){\right)}
\def\delx{\Delta x}
\def\dell{\Delta t}
\def\x{\xi}
\def\dphi{{\partial_{\phi}}}
\def\dphiphi{{\partial^2_{\phi\phi}}}
\def\dxi{{\partial_{\xi}}}
\def\dxixi{{\partial^2_{\xi\xi}}}
\def\dmix{{\partial^2_{\phi\xi}}}
\renewcommand{\nocite}[1]{}
\title{Self-pulsing of electron transmission by a transversal magnetic field} 
\author{Manamohan Prusty} 
\author{Holger Schanz} 
\email{holger@nld.ds.mpg.de}
\affiliation{Max-Planck-Institut f\"ur Dynamik und Selbstorganisation, 
und Fakult{\"a}t f{\"u}r Physik, Universit{\"a}t G{\"o}ttingen,
Bunsenstra{\ss}e 10, D-37073 G\"ottingen, Germany}
\date{\today}
\begin{abstract}
  The distribution of scattering delay times is analyzed for classical
  electrons which are transmitted through a finite waveguide. For non-zero
  magnetic field the distribution shows a regular pattern of maxima
  (logarithmic singularities). Although their location follows from a simple
  commensurability condition, there is no straightforward geometric
  explanation of this self-pulsing effect. Rather it can be understood as a
  time-dependent analog of transverse magnetic focusing, in terms of the
  stationary points of the delay time. We also discuss the possibility of
  singularities in the delay-time distribution for generic 2D scattering
  systems.
\end{abstract}
\maketitle
Ballistic semiconductor nanostructures offer fascinating opportunities to
observe effects from the theory of classical dynamical systems. Recent
experiments with confined two-dimensional electron gases (2DEG) revealed
traces of chaos, nonlinear resonances, unstable periodic orbits or
KAM-hierarchies in phase space \nocite{M+92,W+96,S+98,D+02a}\cite{M+92}, and
also a variety of magnetic commensurability effects
\nocite{vH+89}\nocite{W+91,FGK92}\nocite{F+01d}\cite{vH+89,W+91,F+01d}. Among
the latter, magnetic focusing \cite{vH+89} is of particular importance and has
found numerous applications such as the detection of composite fermions
\cite{GSJ94} or the separation of spin states \cite{R+04}. All these results
were based on conductance
measurements. However, it is well known that the transition probabilities
alone provide only an incomplete description of a scattering system. Important
additional information is contained for example in the delay time, i.e., the
time spent by the scattered particle in the interaction region.  In a seminal
work Wigner had pointed out the equivalence between the delay time and the energy
derivative of the phase of transition amplitudes \cite{Wig55}.  Ever since
there was a lot of theoretical and experimental activity devoted to
understanding the distribution of this quantity in various physical contexts
\cite{dCN02}, including in particular also mesoscopic transport through
systems with nonlinear dynamics
\nocite{FS97}\nocite{G+99,T+99}\nocite{JMS01,J+04a,LJR06}\nocite{D+04}
\cite{FS97,G+99,JMS01,D+04}. For semiconductor nanostructures there exist
pioneering measurements of picosecond delays for ballistic electrons in a
magnetic focusing geometry \cite{SL04}. Moreover, experimental access to the
scattering phase (e.~g., in quantum dots \nocite{Y+95,S+97c,S+04}\cite{Y+95})
might result in an electronic implementation of Wigner's relation in the near
future.

\begin{figure}[!t]
 \centerline{\psfig{figure=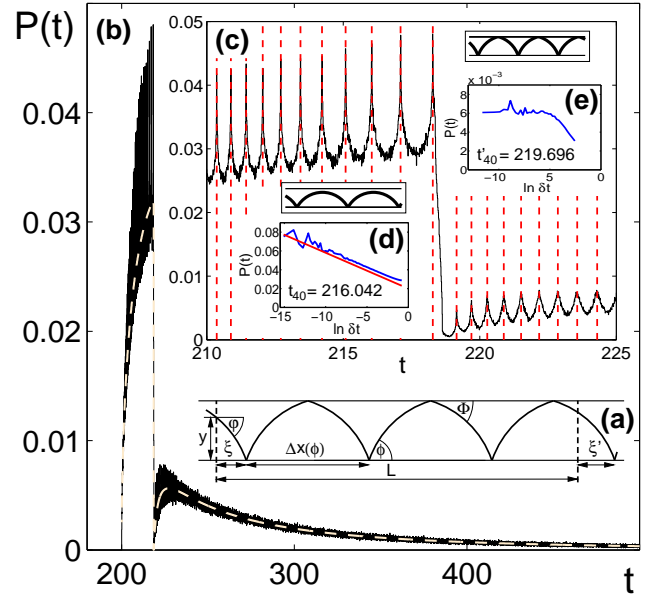,width=85mm}}
 \caption{\label{fig1} (a) A typical electron trajectory transmitting through a
   channel of length $L$ under the influence of a perpendicular magnetic
   field. (b) Distribution of transmission delays for $r=4$ and $L=200$. The
   white dashed line is the average density, Eq.~(\ref{pav}). The inset (c) magnifies
   the region around $t^*\approx 218.5$ (Eq.~(\ref{tstar})). Here the dashed
   vertical lines mark the delay times $t_{n}$ of orbits which are
   commensurate with the channel length (see text). The small insets
   demonstrate with two examples that the maxima $t_{n}<t^*$ (d) actually are
   logarithmic singularities while this is not the case for $t_{n}'>t^*$ (e).
   In these semilogarithmic plots the density is shown as a function of
   $\delta t=t-t_{n}$. The straight line in (d) has the slope given in
   Eq.~(\ref{slope}).}
\end{figure}

Here we analyze the delay-time distribution for the
transmission of electrons through a clean waveguide with a transversal
magnetic field (Fig.~\ref{fig1}a). This model is utterly simple but not
unrealistic in the above context and could be realized, e.~g., as a finite
constriction in a 2DEG. We find a rather surprising self-pulsing effect: A
number of particles entering the interaction region at the same moment of time
exit from the scatterer bunched together in packets which form a nearly
periodic train.  Specifically we calculated numerically the distribution of
trajectory lengths from entry to exit and obtained a histogram which is
composed of well-separated peaks on a smooth background (Fig.~\ref{fig1}b, c)
\endnote{This is not the case in the absence of a magnetic field
  ($r=\infty$). There the distribution is monotonically decreasing,
  $P(t)=L^{-1}(L/t)^3(1-[L/t]^2)^{-1/2}$ for $t>L$.}.

Recently, self-pulsing was predicted in another physical context \cite{JMS01},
and experimental confirmation came from microwave scattering in 2D resonators
\cite{D+04}.  There the scattering echoes revealed a characteristic frequency
of the internal weakly chaotic dynamics. No similar explanation applies to our
(integrable) model. Another striking difference is the fact that in our case
the delay-time density actually diverges at some of the peaks
(Fig.~\ref{fig1}d).

Our numerical results are reminiscent of transverse magnetic focusing
\cite{vH+89}, where the conductance between two point contacts to a 2DEG
oscillates as a function of the magnetic field (or some other parameter) and
shows a regular pattern of singularities \endnote{In reality these
  singularities are blurred by the finite contact width or by quantum effects
  \cite{vH+89}. An analogous smoothing of the delay-time
  singularities must be expected, but this will not be considered here.}. The
fundamental dynamical origin of these singularities are caustics of the
classical electron motion.  They correspond to the stationary points of a
function that maps the incident angle to the exit location for an electron
beam. Similarly we will explain the self-pulsing by stationary points of the
delay time in the asymptotic phase space of scattering trajectories. To our
knowledge, the effect of such stationary points has never been analyzed
before, not in any kind of scattering system. Therefore our results might be
of interest beyond the study of electrons in a magnetic field, just as
caustics in configuration space are relevant beyond magnetic focusing and in
many different physical systems
\nocite{Wri86,BW94,T+01,Kap02,WMB06}\cite{Wri86}. We shall come back to this
point at the end of the paper.

We consider electrons entering the channel of Fig.~\ref{fig1}a from the left
and with a given Fermi energy $E_{F}=(m/2)v_{F}^2$. Otherwise the initial
conditions $0\le y\le 1$ and $|\varphi|\le+{\pi\/2}$ are randomly drawn from
the microcanonical distribution restricted to $x=0$,
$p(y,\varphi)={1\/2}\cos\varphi$. We use the channel width and the Fermi
velocity as units of length and velocity, respectively ($v_{F}=b=1$).  The
cyclotron radius $r=mv_{F}/eB$ and the length $L$ of the system are free
parameters \endnote{Our results are not sensitive to the precise values
  of $L$ and $r$ as long as the magnetic field has a moderate strength
  $r\gtrsim 1$.  For numerical calculations we choose $r=4$ and $L=200$.}.  Any
trajectory is composed of circular arcs which are traversed clockwise (by
convention about the direction of the magnetic field). For a point
$(x,y,\varphi)$ in phase space, the center of the current arc is the point
$(x_{c},y_{c})=(x+r\sin\varphi,y-r\cos\varphi)$. Upon reflection from the
channel walls $y_{c}$ is conserved.  Thus the system is integrable. We
restrict attention to $y_{c}<0$.  For these trajectories the $x$-component of
the velocity is positive at any moment of time, and they generate the bulk of
the left-to-right transmission delay distribution (up to $t\sim 1270$ for the
parameters of Fig.~\ref{fig1}). We change variables to $(\xi,\phi)$, where
$0\le\phi=\arccos(-y_{c}/r)\le\pi/2$ is the direction of the trajectory
immediately after a reflection from the lower wall. In terms of this angle,
the average transport velocity of a trajectory in an infinite channel is
$\overline v_{x}={\delx(\phi)/\dell(\phi)}$.  Here
\begin{eqnarray}
\label{delx}
\delx(\phi)&=&2r(\sin\phi-\sin\Phi)\,,
\\
\label{dell}
\dell(\phi)&=&2r\,(\phi-\Phi)
\end{eqnarray}
are the horizontal and the total length of the trajectory segment between two
reflections from the lower wall, respectively. If $\cos\phi\le 1-1/r$, the
trajectory reaches the upper wall and is reflected there with the angle
$\Phi=\arccos(\cos\phi+{1\/r})$ (see Fig.~\ref{fig1}a). $\Phi$ is set to zero
for trajectories skipping along the lower wall only.  $\xi$ is the
longitudinal location of the first reflection from the lower wall.  We make
the convention $-\delx(\phi)/2\le\xi<\delx(\phi)/2$ such that negative
values of $\xi$ correspond to positive $\varphi$ and vice
versa. Explicitly, the transformation between $(\varphi,y)$ and $(\phi,\xi)$ is
given by
\begin{eqnarray}
\label{y}
y&=&r(\cos\varphi-\cos\phi)\,,
\\
\label{xi1}
-2\x&=&2r(\sin\phi-\sin\varphi)
\qquad(\xi<0)\,,
\\
\label{xi2}
+2\x&=&2r(\sin\phi+\sin\varphi)
\qquad(\xi>0)\,.
\end{eqnarray}
The probability density in the new variables is $p(\phi,\xi)={1\/2}\sin\phi$.

As a first approximation to the delay time we ignore the dependence on
$\xi$ and set
\begin{equation}\label{tav}
\overline\tau(\phi)={L\/\overline v_{x}}=L{\dell(\phi)\/\delx(\phi)}\,.
\end{equation}
This results in the distribution
\begin{eqnarray}\label{pav}
\overline P(t)&=&\int d\phi\,d\xi\,p(\phi,\xi)\,\delta(t-\overline\tau(\phi))
\nonumber\\&=&{\sin\phi_{t}\,\delx(\phi_{t})\/2|\overline\tau'(\phi_{t})|}\,,
\end{eqnarray}
where $\phi_{t}$ is the root of $\overline\tau(\phi)=t$. For orbits which do
not reach the upper wall we find
\begin{eqnarray}\label{dtav}
\overline\tau'(\phi)
&=&L{\sin\phi-\phi\cos\phi\/\sin^2\phi}\qquad(\cos\phi>1-r^{-1})\,,
\\
\label{pav1}
\overline P(t)&=&{r\/L}{\sin^4\phi_{t}\/\sin\phi_{t}-\phi_{t}\cos\phi_{t}}
\qquad\(L<t<t^{*}\)\,.
\end{eqnarray}
Clearly, the transmission delay is bounded from below by $t=L$. This
corresponds to trajectories skipping along the lower wall with an
infinitesimal reflection angle $\phi\to 0$ such that their velocity is always
directed along the channel. Close to the minimum we have according to
Eq.~(\ref{tav}) $t\sim L(1+\phi_{t}^2/6)$ and from Eq.~(\ref{pav1}) $\overline
P(t)\sim 3r\phi_{t}/L$. This explains the square-root like onset of the
density, $P(t)\propto\sqrt{t-L}$, which is observed in Fig.~\ref{fig1}b at
$t=200$.  On the other hand the time
\begin{equation}\label{tstar}
t^{*}(L,r)=L{\arccos(1-r^{-1})\/\sqrt{2r^{-1}-r^{-2}}}
\end{equation}
is the maximum delay for skipping orbits. Their contribution to $\overline
P(t)$ abruptly drops to zero at this point ($t^{*}\approx218.5$ in
Fig.~\ref{fig1}b). Higher values $t>t^{*}$ correspond to trajectories which
bounce off both, the lower and the upper wall of the waveguide. We omit here the
lengthy expression replacing Eq.~(\ref{dtav}) in this case.
After substitution into Eq.~(\ref{pav}) it yields the averaged delay-time
distribution for $t>t^{*}$ which is displayed together with Eq.~(\ref{pav1}) by the
dashed white line in Fig.~\ref{fig1}b.

Now we come to the conspicuous oscillations in the delay-time density around
its average value (Fig.~\ref{fig1}c). In general the delay time
$\tau(\phi,\xi)$ depends on both variables, $\phi$ and $\xi$. A variation of
$\xi$ corresponds to a longitudinal shift of the trajectory (Fig.~\ref{fig1}a)
and results in \endnote{We use a notation where, e.~g., $\dxi\tau$
  stands for $\partial\tau/\partial\xi$. All second-order derivatives are
  evaluated at the stationary point $(\phi,\xi)=(\phi_{n},0)$. }
\begin{equation}\label{deriv}
{{\dxi}\tau(\phi,\xi)}=(\cos\varphi)^{-1}-(\cos\varphi')^{-1}\,,
\end{equation}
where $\varphi$ and $\varphi'$ are the inclination angles of the trajectory on
entry and exit, respectively. However, if the
length of the system is an integer multiple of the length of a segment,
\begin{equation}\label{cond}
L=n\,\delx(\phi_{n})\,,
\end{equation}
these angles are equal. Only in this case the dependence of the delay time on
$\xi$ disappears and we have
\begin{equation}\label{noxidep}
\tau(\phi_{n},\xi)=\overline\tau(\phi_{n})=n\dell(\phi_{n})\,.  
\end{equation}
For each $n$
there can be at most two solutions to Eq.~(\ref{cond}), one corresponding to a
skipping orbit and the other one not.  We denote the corresponding delay times
by $t_{n}<t^{*}$ and $t_{n}'>t^{*}$, respectively. In Fig.~\ref{fig1}c these
values are marked with dashed vertical lines, and we see that they coincide
with the peaks in the delay-time distribution.  It is tempting to explain this
observation by the following simple (and misleading) argument: Typically, for
a given $\phi$ the various values of $\xi$ contribute at different times to
the density, while all of them add up in the same bin when $\phi=\phi_{n}$. So
it is clear that the count should have a peak there!  However, this cannot
explain why the peaks at $t_{n}$ are singularities of the density
while the $t_{n}'$ are finite maxima.  To highlight this fact numerically we
have magnified in Fig.~\ref{fig1}d the density in the vicinity of $t_{40}$ and
observe that it grows as $P(t)\propto\ln(t-t_{n})$.  In contrast the
corresponding plot for $t_{40}'$ (Fig.~\ref{fig1}e) saturates to a
finite value.

\begin{figure}[htb]
 \centerline{\psfig{figure=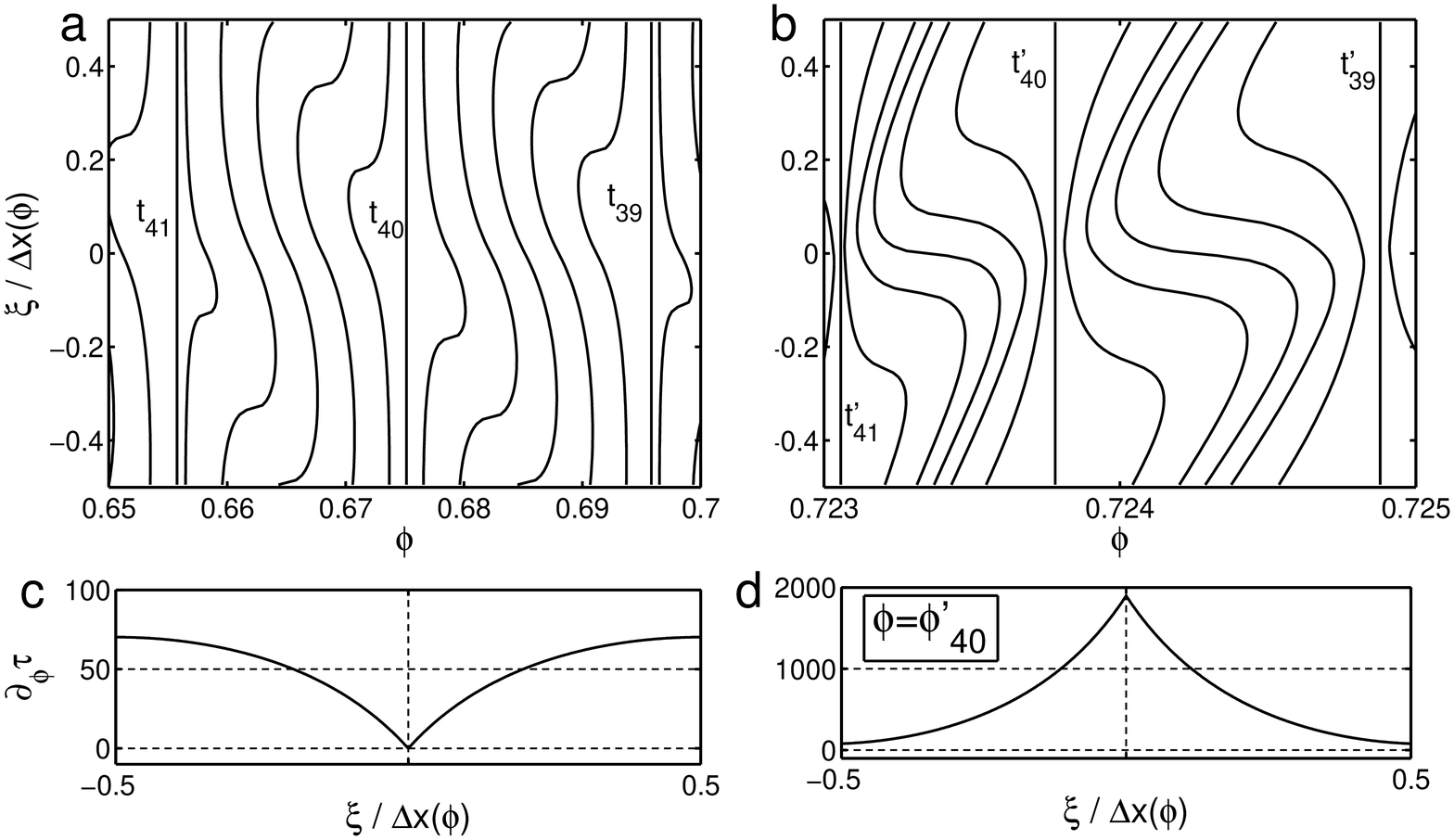,width=85mm}}
 \caption{\label{fig2} Lines of constant delay time $\tau(\xi,\phi)=t$
   are shown in the vicinity of (a) $t_{40}$ and (b) $t'_{40}$. The function
   ${\dphi}\tau$ vanishes at $\phi=\phi_{n}$ and $\xi=0$ (c) while no
   stationary point exists for $\phi=\phi'_{n}$ (d).}
\end{figure}

In order to understand the nature of the maxima we construct the function
$\tau(\phi,\xi)$ explicitly and represent the delay-time density as
\begin{eqnarray}
P(t)&=&\int d\phi\,\int d\xi\, p(\phi,\xi)\,\delta(t-\tau(\phi,\xi))
\nonumber\\&=&\left.\int d\xi {p(\phi,\xi)\/|{\dphi}\,\tau|}\right|_{\phi=\phi(t,\xi)}\,.
\label{pint}
\end{eqnarray}
To this end we decompose a trajectory into a number $n$ of complete arcs and
two additional terms for entry and exit. Formally we define $n(\phi)$ as the
best integer approximation to $L/\delx$, e.~g.\ $n=3$ in Fig.~\ref{fig1}a.
The mismatch to exact commensurability will be denoted by
\begin{equation}
\delta L(\phi)=n\delx(\phi)-L\,.
\end{equation}
According to Fig.~\ref{fig1}a the delay time is given by
\begin{eqnarray}\label{tau}
\tau(\phi,\x)&=&n\,\dell(\phi)+s(\phi,\x)-s(\phi,\xi')
\end{eqnarray}
with $\xi'=\xi+\delta L(\phi)$.
In Eq.~(\ref{tau}), $s(\phi,\xi)$ is the length of the incomplete arc segment
at the beginning of the trajectory. For $\xi$ in Fig.~\ref{fig1}a this is
\begin{eqnarray}\label{s1}
s(\phi,\x)&=&r(\phi+\varphi)\qquad\qquad\quad(\varphi<0,\ 0\le\xi\le {\delx(\phi)/2})
\nonumber\\&=&r[\phi+\arcsin(\x/r-\sin\phi)]
\end{eqnarray}
(the second line follows from Eq.~(\ref{xi2})). However, $\xi'$
can take values in the interval $[-\delx(\phi),\delx(\phi)]$ since we
have $|\delta L(\phi)|<{\delx\/2}$ and $|\xi|<{\delx\/2}$. Therefore we extend
the domain of $s(\phi,\xi)$  by the definitions
\begin{eqnarray}
\label{cont1}
s(\phi,-\xi)&=&-s(\phi,\xi)\,,
\\
\label{cont2}
s(\phi,\delx(\phi)-\xi)&=&\dell(\phi)-s(\phi,\xi)\,.
\end{eqnarray}
Note that $s(\phi,\xi)$ is continuous at $s(\phi,0)=0$ and
$s(\phi,\delx(\phi)/2)=\dell(\phi)/2$ but not analytic at $\xi=0$.
Eqs.~(\ref{tau})-(\ref{cont2}) represent the delay time for
$0\le\phi\le{\pi\/2}$ and arbitrary values $-{\delx\/2}<\xi<+{\delx\/2}$.
Fig.~\ref{fig2} displays this function in the regions (a) $t_{41}\lesssim
t\lesssim t_{39}$ and (b) $t'_{41}\lesssim t\lesssim t'_{39}$. All level sets
of constant delay time are one-dimensional curves in the two-dimensional phase
space. In this respect there is nothing special about $t=t_{n}$.  Rather we
observe in Fig.~\ref{fig2}c ${\dphi}\tau(\phi_{n},0)=0$. This is not
immediately obvious from geometrical considerations but it can be confirmed
analytically. We conclude that $(\phi,\xi)=(\phi_n,0)$ are stationary points
of the delay time, and this is what really singles out the values $t=t_n$. No
stationary points exist for $t>t^{*}$ (Fig.~\ref{fig2}d), although $\dphi\tau$
is very small close to $\xi=\pm\delx(\phi)/2$.

In the vicinity of the stationary points we can expand the delay time to second
order,
\begin{eqnarray}\label{tauapp}
\delta t={1\/2}{\dphiphi}\tau\,\delta^2\phi
+{\dmix}\tau\,\xi\,\delta\phi\,
+{1\/2}{\dxixi}\tau\,\xi^2\,,
\end{eqnarray}
solve for $\delta\phi=\phi-\phi_{n}$ and substitute the result into
\begin{eqnarray}\label{dtauapp}
\partial_{\phi}\tau&=&{\dphiphi}\tau\,\delta\phi+{\dmix}\tau\,\xi\,.
\end{eqnarray}
This gives $\dphi\tau$ as a function of $\xi$ and $\delta t=t-t_{n}$ and can
be used in Eq.~(\ref{pint}) to calculate the divergent contribution
to the delay-time density. Unfortunately this procedure is more complicated
than it may seem on first sight. Due to Eq.~(\ref{cont1}) we must distinguish
for $\delta t,\delta L>0$ three regions (i) $\xi>0$, (ii) $\xi_{0}<\xi<0$ and
(iii) $\xi<\xi_{0}$ (and again three regions for $\delta t,\delta L<0$). Here
$\xi_{0}(\delta t)$ is the point where the third term in Eq.~(\ref{tau})
changes sign. It is given implicitly by the condition $\xi_{0}=-\delta
L(\phi_{n}+\delta\phi)$ where $\delta \phi(\xi_{0},\delta t)$ is the root of
Eq.~(\ref{tauapp}). Solving for $\xi_{0}$ we find
\begin{equation}\label{xi0}   
\xi_{0}=-(\sin\phi_{n})^{-1}\sqrt{2\delta t\,(\cos^3\phi_{n})/(2n-1)}\,.
\end{equation}
Note that we did not drop the third term in Eq.~(\ref{tauapp}). In fact we
have $\dxixi\tau\ne 0$ in the second region although $\tau(\phi_{n},\xi)$ is
constant as a function of $\xi$. This is no contradiction since, according to
Eq.~(\ref{xi0}), this region shrinks to a point for $\phi=\phi_n$ (and hence is
not visible in Fig.~\ref{fig2}d). With Eqs.~(\ref{tauapp})-(\ref{xi0}) it can
be shown that the contribution of region (ii) approaches a constant as $\delta
t\to 0$. Hence it is irrelevant for our purpose. In the regions (i), (iii) we
use $\dxixi\tau=0$ and find
\begin{eqnarray}\label{indefint}
\dphi\tau(\xi,\delta t)&=&\sqrt{(\dmix\tau)^2\,\xi^2+2\delta t\,\dphiphi\tau}\,,
\\
\int{d\xi\/|\dphi\tau|}&=&{1\/|\dmix\tau|}\ln\(|\dmix\tau|\xi+|\dphi\tau|\)+\mbox{const.}\quad\mbox{}
\end{eqnarray}
In region (i) the integral extends from $\xi=0$ to a value
$\sim\delx(\phi_{n})/2$. Its leading contribution is the logarithmic term
$|\ln\delta t/2\dmix\tau|$ from $\xi=0$.  The same contribution
results from the upper limit $\xi_{0}$ of region (iii), and in either case we
have $|\dmix\tau|=(L/r)(\cos\phi_{n})^{-2}$. Thus, together with the
prefactor ${1\/2}\sin\phi_{n}=L/4nr$ from Eq.~(\ref{pint}), we get
\begin{equation}\label{slope}
P(t)={1-(L/2nr)^2\/4n}|\ln(t-t_{n})| \qquad(t\to t_{n})\,.
\end{equation}
This prediction for the strength of the divergence has been confirmed
 numerically. For example both, a fit to the data in
Fig.~\ref{fig1}d and Eq.~(\ref{slope}) yield a
prefactor $0.00381$ for $n=40$. 

The rather complicated analytical structure in the vicinity of the stationary
points of the delay time, and also the self-pulsing due to their regular
arrangement, are specific to our model. We would like to stress, however, that
these features are not at all necessary for the appearance of singularities in
the delay-time density. Generic stationary points for 2D scattering problems
are extrema and saddle points. It is easy to see that saddle points lead again
to a logarithmic divergence, while the density is finite at isolated maxima
and minima.  Although a logarithm is a rather weak singularity, it should lead
to a prominent maximum in a histogram with finite resolution such as
Fig.~\ref{fig1}c. This might be important, e.~g., for inverse scattering
problems. Consider for example a situation where it is difficult to control
the precise initial conditions of test particles while their time delay can be
measured with high precision.  Then saddle points of the delay time will be
among the dynamical features which are directly accessible from experimental data.
Therefore it seems worthwhile to study them in more generic situations and
independent of self-pulsing effects.

\end{document}